*

# Experimental Analysis of a Self-Coherent M-QAM Receiver by Means of Recurrent Optical Spectrum Slicing and Direct Detection

Kostas Sozos, Francesco Da Ros, *Senior Member Optica*, Metodi Yankov, Stavros Deligiannidis, George Sarantoglou, Charis Mesaritakis and Adonis Bogris, *Fellow Optica*

***Abstract*— High order modulation formats constitute the most prominent way for increasing spectral efficiency in transmission systems. Coherent transceivers that support such higher order formats require heavy digital signal processing (DSP), which increases the power consumption of coherent pluggables, well above the intensity modulation and direct detection (IM/DD) counterparts. Self-coherent or phase retrieval methods have emerged as potential solutions, trying to combine the merits of coherent technology with the simplicity of direct detection. In this work, we experimentally demonstrate the reception of quadrature amplitude modulation (QAM) modulation formats based on direct detection aided by the recurrent optical spectrum slicing (ROSS) photonic accelerator, utilizing minimal DSP and low modulator driving voltages. We experimentally demonstrate 32 Gbaud QAM-4/16 for 25 km, 50 km and 75 km in the C-band aided by a linear digital equalization and the use of programmable photonics as recurrent optical spectrum slicers. We showcase successful detection with driving swings below Vπ/3 in contrast to the full swing required by conventional coherent transceivers. We further improve the system performance utilizing geometric constellation shaping. Finally, we explore the potential power consumption improvement for the next-generation 1.6T pluggables, showcasing over 40% reduction with respect to the most lightweight state of the art coherent solutions reported in literature.**

## I. INTRODUCTION

FIELD recovery of both the amplitude and the phase of optical signals revolutionized modern communications, offering increased spectral efficiency and full compensation of the transmission impairments in the digital domain. Today, coherent technology tends to dominate over legacy intensity modulation and direct detection (IM/DD) systems even in the data center interconnection environment [1]. However, this transition to coherent solutions for short-reach applications, translates to increased power consumption that challenges the thermal limits of the OSFP and even CFP8 form factor pluggables [2]. Coherent detection requires heavy digital signal processing (DSP) for discriminating the two polarizations, equalizing the signal and recovering the phase. On top of that, baudrate sampling and equalization is yet to be achieved in practical coherent systems, requiring higher bandwidth, resolution and power consumption for the ADCs [3], [4].

In the last decade, plenty of solutions have been proposed, trying to bridge the gap between the benefits of higher spectral efficiency enabled by digital equalization and the drawbacks of the complicated, expensive and energy-hungry coherent transmission. The most important of them, the Stokes Vector Field Recovery Receivers [6] and the Kramers-Kronig (KK) receivers [7], have not found a clear path towards commercialization yet, with the most prominent solutions for 800G LR1 eventually being based on the typical coherent receiver through the so-called coherent-lite approaches [1], [8]. Although all the competing techniques are elegant solutions, they are characterized by specific deficiencies which have hindered their practical deployment up to now. KK algorithm broadens the signal spectrum considerably, necessitating a higher sampling rate, at least in the Nyquist rate or two times the baudrate (2 samples per symbol) [9], thus increasing the power consumption of the ADCs and the equalizers. Stokes-vector receivers at 3D variations [10], combined with the KK algorithm and single sideband signals in one polarization, achieve up to 75% of the electrical spectral efficiency of a coherent transceiver with a slightly simpler receiver architecture, but with no evident gain in terms of DSP simplification or power savings. Self-homodyne receivers also relax the linewidth requirements of the laser and avoid carrier phase estimation algorithms, but necessitate full-duplex fiber

* The work has received funding from the Hellenic Foundation for Research and Innovation (H.F.R.I., Project Number: 2901), from the EU Horizon Europe PROMETHEUS project (101070195) and the Villum foundation (OPTIC-AI grant n. VIL29344). *(Corresponding author: K. Sozos).*

K. Sozos, S. Deligiannidis and A. Bogris are with the University of West Attica, Dept. of Informatics and Computer Engineering. Aghiou Spiridonos, 12243, Egaleo, Athens, Greece, ksozos@uniwa.gr

C. Mesaritakis and G. Sarantoglou are with the University of West Attica, Dept. of Biomedical Engineering. Aghiou Spiridonos, 12243, Egaleo, Athens, Greece, cmesar@aegean.gr

Francesco Da Ros, Metodi P. Yankov are with the Dept. of Electrical and Photonics Engineering, Technical University of Denmark, DK-2800 Kgs. Lyngby, Denmark, {fdro, meya}@dtu.dk

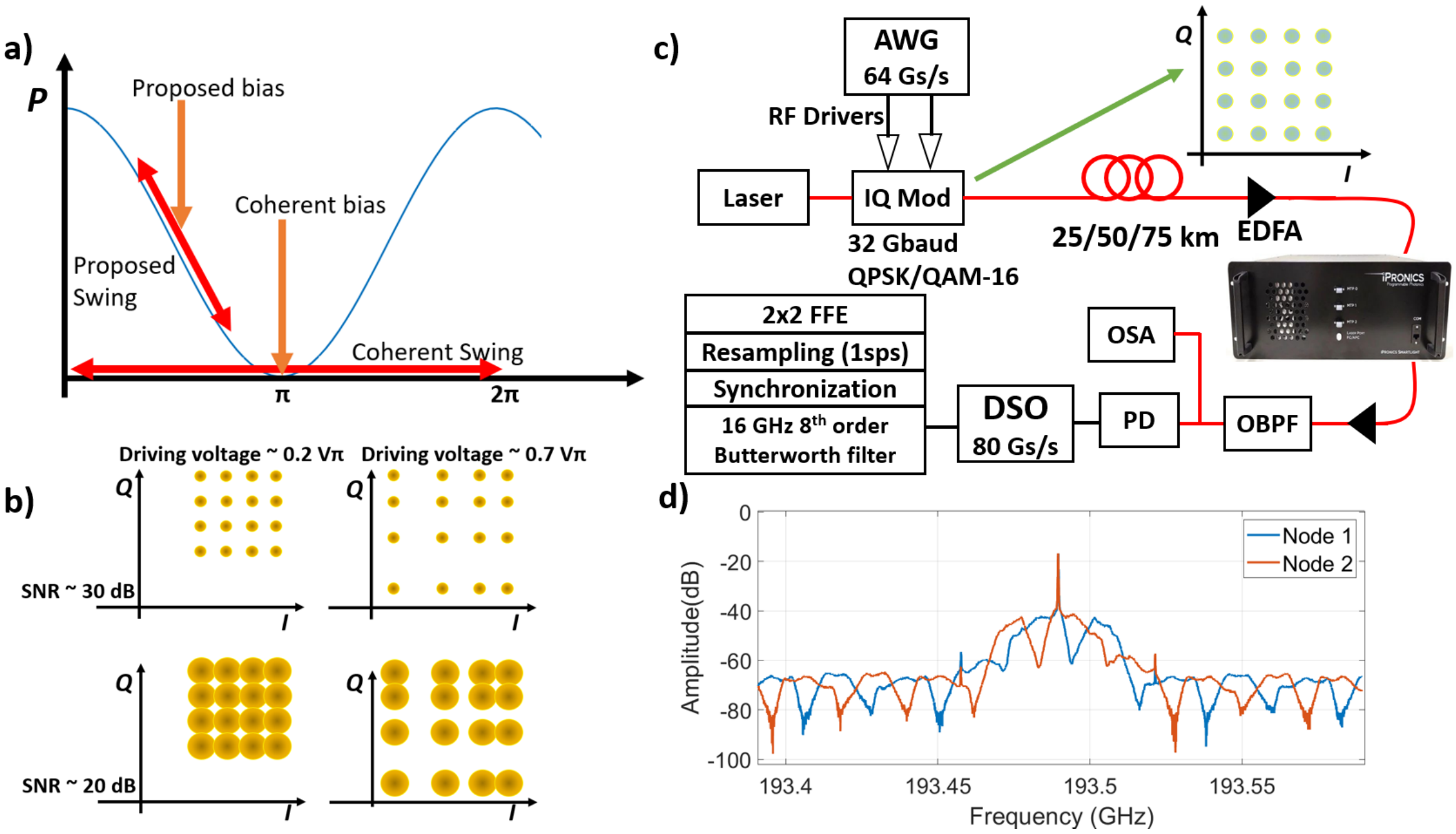

Figure 1. The concept of the ROSS coherent receiver. a) The IQ modulator bias condition of the proposed transmitter in comparison with the typical coherent transmitter. b) The constellation that stems from the proposed modulator in different Driving voltage and SNR regimes. c) The experimental setup that is used in this work, with the Ipronics Smartlight processor implementing the ROSS nodes, followed by an optical band-pass filter (OBPF) a photodetector (PD), an optical spectrum analyzer (OSA) and the digital signal oscilloscope (DSO). d) The optical spectrum of the received signal after two indicative ROSS nodes, as measured by the OSA.

pairs for the transmission of the carrier in parallel with the signal, imposing extra losses [11]. Phase extracting methods eliminate the local oscillator, the optical hybrid, the need for balanced detection and exhibit the same optical spectral efficiency as a coherent receiver [12]. Yet, the electrical spectral efficiency remains lower due to the wider receiver bandwidth and the digital upsampling that is required. Moreover, such methods are followed by latency and large data overhead [13], which adds on top of the heavy forward error correction (FEC) overhead required. Finally, asymmetric self-coherent detection [14] improves the electrical spectral efficiency with respect to other self-coherent proposals, however, the reconstruction of the complex signal is made in the frequency domain, which requires a fast Fourier transform.

In order to maximize the use of the available analog bandwidth, it is beneficial to utilize optical signal processing before the receiver, offloading part of the digital equalization and using low bandwidth components at baudrate sampling. Towards this path, we have recently proposed recurrent optical spectrum slicing (ROSS) neuromorphic receivers for the mitigation of colored noise effects in IM/DD links in the optical domain [15]. ROSS receivers can effectively linearize the channel by optically equalizing the frequency/phase dependent impairments and at the same time provide nonlinear phase to intensity conversion so as to map all constellation points in the amplitude domain. In previous works, we numerically studied a coherent ROSS neuromorphic scheme for simultaneous phase extraction/demodulation and equalization of in phase and quadrature (IQ) modulated signals with the use of two or more recurrent optical filter nodes [16].

Here, we experimentally validate this proposed method taking advantage of a reconfigurable photonic platform to implement the ROSS nodes. In order to enhance the process of symbol discrimination through the conversion of the phase information to the amplitude domain, we contain the modulation of the IQ signal in the first quadrant, by biasing the modulators in the quadrature point and applying driving swing lower than $V_{\pi}$. Preserving the carrier in that way, we avoid the multiple drawbacks of other carrier insertion methods [17]. In order to combat overall system nonlinearities while exploiting the maximum possible modulation amplitude, we apply geometric constellation shaping (GCS) through a process under which we adjust the constellation for the nonlinear modulation and the specific noise distribution of our experimental setup. We study the capabilities of the proposed system in short-reach dispersive channels at 32 Gbaud and up to 75 km, in C-band. By employing two or four single-photodetector receivers followed either by a simple feed-forward equalization (FFE) or light machine learning algorithms, we achieve bit-error ratio (BER) performance in the order of $10^{-5}$ for quadrature phase shift keying (QPSK) and in the order of $10^{-3}$ for quadrature amplitude modulation (QAM)-16 formats in back-to-back, proving its high flexibility in supporting multiple M-QAM formats. Considering a FEC with 15.31% overhead and threshold bit error rate (BER) of $2\times10^{-2}$, QAM-16 reception and equalization at a net bit rate >110 Gb/s is successfully demonstrated for a transmission distance up to 50 km in the C-band. Hence, the same receiver can be flexibly adapted to a wide range of modulation formats and transmission distances in the short reach framework. A detailed power consumption analysis indicates that the specific receiver may offer a 40% power consumption gain if implemented in a custom designed chip compared to lighweight coherent solutions currently considered in short reach applications.

## II. Recurrent Optical Spectral Slicing Self-Coherent Receiver

### *A. Concept and filtering scheme*

Fig. 1 describes the idea behind the proposed transceiver. Instead of biasing the two branches of the IQ modulator in the null point, as in typical coherent transmission, we choose to bias in the quadrature point, as typical for IM/DD systems. Also, instead of applying a driving voltage of two times the Vπ of the modulator, we contain it in values below Vπ (Fig 1a). In this way, the signal remains within one quadrant and the carrier is partially preserved for efficient direct detection, as in various self-coherent implementations, like the KK receiver, which require the transmission of a minimum phase signal. When the driving swing approaches $V_{\pi}$, the constellation becomes nonlinear, while by reducing the swing the signal-to-noise ratio (SNR) deteriorates and the symbol distance decreases as shown in Fig. 1b. In the receiver side, two or more ROSS nodes spectrally slice the signal by means of offset optical filtering with their central frequencies slightly detuned (in the order of 5 GHz for a 32 Gbaud signal) with respect to the signal carrier. This detuned recurrent filtering acts as a neuromorphic engine twofold. On the one hand it applies different weights in different frequencies of the signal and on the other hand it nonlinearly transforms the phase information into the amplitude domain [15]. The nonlinearity of the process is controlled through the recurrent component of the node in terms of the transfer function of each slicer (see fig. 1d). The nonlinear phase to intensity conversion offered by the recurrent spectral slicers assists the receiver to recover the IQ information entirely at the amplitude domain with simple photodetection and linear post processing utilizing one FFE per quadrature. At the same time, colored noise effects like the limited bandwidth of the components and the power fading effect due to fiber dispersion are mitigated through the frequency diversity of the multiple nodes [15]. The mathematical expression of the recurrent filters can be found in [16].

In order to implement the ROSS nodes, we exploit the reconfigurable photonic Smartlight processor [19]. Smartlight contains a photonic mesh consisting of tunable phase shifters or programmable unit cells (PUCs) and can implement various elements from simple optical filters to more complex structures. However, this reconfigurability comes to the price of high losses, such as insertion losses in the order of 0.5 dB at each PUC and over 15 dB for the connection to/from the photonic mesh, leading to SNR limitations in this experiment. These SNR limitations is the only reason that higher order formats, such as QAM-32/64 are not examined in this work.

Through such photonic mesh, we implement a recurrent MZDI with a feedback loop, as shown in Fig. 2a and analytically described in the next section. This recurrent filter node exhibits frequency tunability through a phase shifter in the one arm of the MZDI. The phase difference between the two arms of the MZDI, $\Delta\varphi$, corresponds to a frequency shift of 11 GHz per π.

### *B. Experimental Setup*

The experimental setup that is used in this work, is depicted in Fig. 1c. At the transmitter side, a pseudorandom unrepeated QAM-4/16 sequence is produced in Matlab. The QAM data are uploaded to the two channels of an arbitrary waveform generator (AWG) with 23 GHz analog bandwidth at 2 samples per symbol (sps) corresponding to the I and Q complex signal components. The resulted signal is resampled to the sampling rate of AWG operating at 64 Gsa/s. Two 55 GHz radio frequency (RF) drivers amplify the I and Q electrical signals by 14 dB and a 25-GHz IQ modulator with Vπ=3.5 V is used to modulate the signal. The modulator bias is set at the quadrature point in order to constrain the QAM constellation to the one I-Q quadrant, while the AWG output voltage is varied from 150 mV to 600 mV which translates to values between 0.2Vπ and 0.85Vπ of the modulator after the RF amplifier. The modulated carrier is produced by a tunable C-band source operating at 1549.4 nm with 10 dBm of output power. The optical signal, with approximately 3 dBm launched power after the MZM, is propagated through different lengths of single-mode fiber (SMF) and then is amplified by an erbium-doped fiber amplifier (EDFA). In this work, we study transmission at 25 km, 50 km and 75 km spools with dispersion coefficient D=16.4 ps/nm/km.

The ROSS nodes follow after the amplification implemented with the use of the iPronics Smartlight photonic processor. The minimal overall Smartlight losses are 22.5 dB, necessitating the use of a second EDFA followed by an extra 1-nm wide optical filter to remove out-of-band noise. Then, a 99-1 coupler feeds an optical spectrum analyzer (OSA, low-power tap) and a photodetector of 50 GHz bandwidth. Finally, an 80 GSa/s digital signal oscilloscope (DSO) is utilized, capturing 10 instances of the received data series for each filter under test which are sequentially measured. The receiver-side offline processing includes digital low-pass filtering at 16 GHz with a $8^{th}$ order Butterworth filter, framing synchronization carried out separately for each filter position and downsampling to 1 sps as we want to focus on the lowest complexity and cost-efficient symbol processing approach which makes it compatible to very high baud rate signals (>100 GBaud). The selected synchronized and resampled outputs are provided as inputs to a symbol-spaced FFE with 21 taps for each filter and each quadrature, targeting separately the real and the imaginary part of the signal which are then composed to reconstruct the full complex symbols. This corresponds to a real valued Nx2 MIMO FFE, with N the number of the employed filter nodes. The equalizer length was chosen after an optimization process revealing performance saturation for a larger number of taps. We launch up to 4 filter outputs in the FFE post-processor, therefore the final weights scale from 84 (2 filter outputs) to 168 (4 filter outputs). We use 10000 symbols for training via a linear regression algorithm. In the QAM demodulation process, BER is calculated by direct error counting. We use 45000 symbols for testing and we average the BER obtained in five DSO

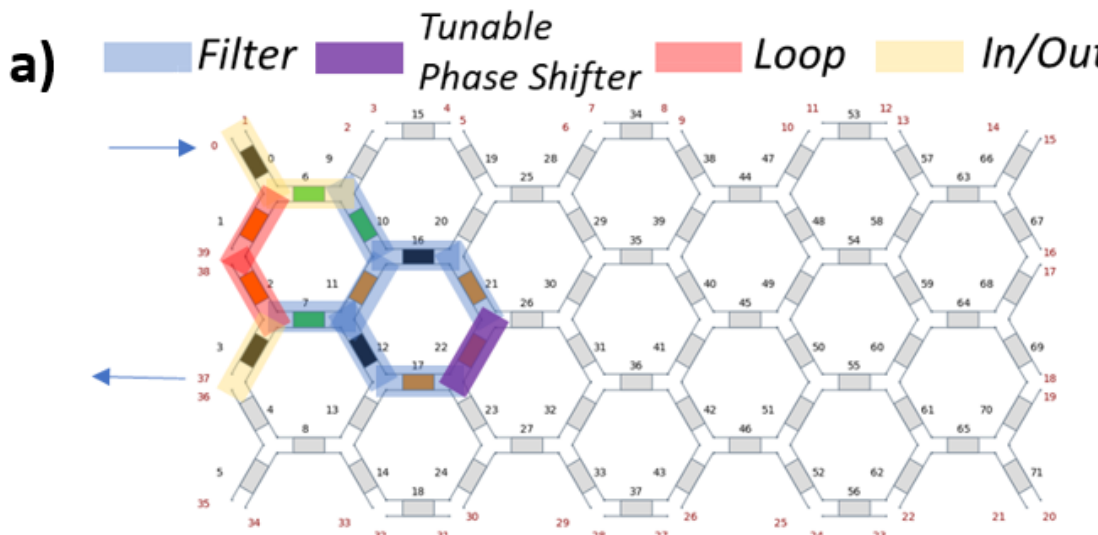

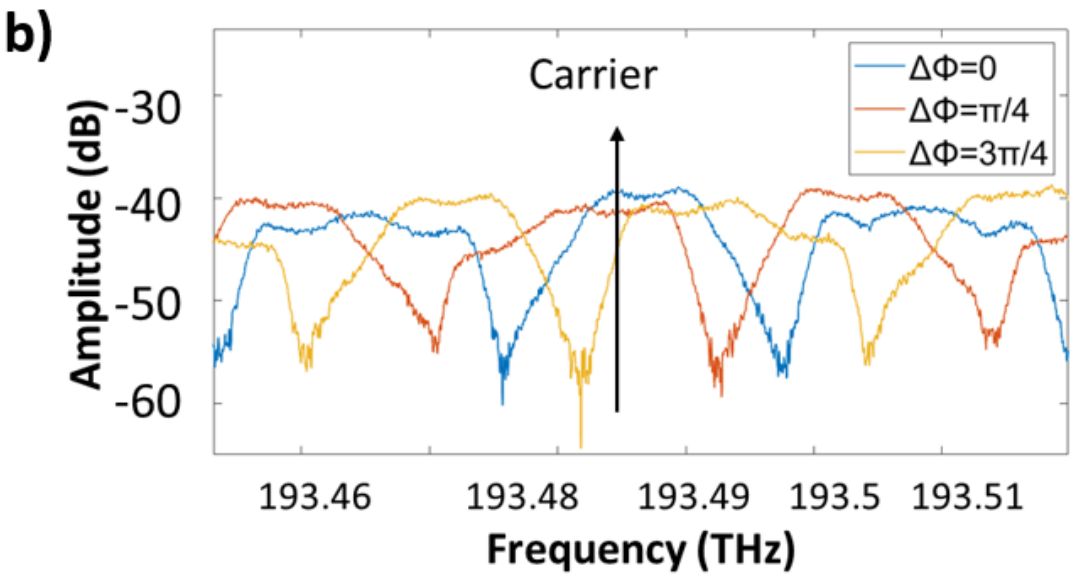

Figure 2. The characteristics of the ROSS node implementation in the Smartlight Processor. a) The configuration of the recurrent MZDI employing 12 PUCs. b) The spectral response of the node for three different $\Delta\varphi$ values, corresponding to different frequency offsets from the carrier.

captures.

The ROSS nodes in this experiment are demonstrated through the iPronics Smartlight processor [19]. We utilize the programmable unit cells (PUCs) of the processor in order to create a MZDI with 4 PUCs path difference ($\Delta T$=45 ps) as the optical filter and a feedback loop of 2 PUCs, with length of 1.622 mm and delay of 22.5 ps. The path difference in the MZDI determines the free spectral range (FSR) of the filter and the subsequent bandwidth, which are 22.2 GHz and 11.1 GHz respectively. The node configuration is depicted on top of the Smartlight hexagonal mesh structure in Fig. 2a. The node frequency response depends on the tunable phase shifter, shown in the Fig. 2a with purple color, by applying a phase shift, $\Delta\varphi$, between the two paths of the MZDI. The spectral response of the MZDI node for different $\Delta\varphi$ values is depicted in Fig. 2b. It is shown that $\Delta\varphi$ leads to a frequency shift, while also affecting the transfer function shape. The loss breakdown of this implementation is shown in the table I. It is shown that the two-PUC feedback loop attenuates the signal by about 1 dB, while alternative node implementations could increase the feedback losses to 2 dB and even 6-7 dB. Increasing feedback losses would diminish the extinction ratio of the secondary peaks in the transfer function of the recurrent node. One can easily conclude that most of the loss sources in Table I could be eliminated exploiting a dedicated photonic integrated implementation. For example, the input/output losses including input-output waveguides could be substantially reduced through better packaging and photodetection units co-integrated with the photonic processor. Moreover, the recurrent MZDI could achieve attenuation in the order of 3 dB from the 6.5 dB of the Smartlight implementation. On the other hand, extra attenuation of 3 (6) dB for splitting should be added in an

TABLE I

*Loss breakdown for each stage of the recurrent MZDI node implemented with the Smartlight processor.*

| Loss Source | Loss Analysis | Total Loss |
|---|---|---|
| **Single PUC losses** | 0.5 dB | |
| **Input/Output waveguide Connection/coupling to the mesh** | 2 x 7-8 dB | 14-16 dB |
| **MZDI PUCs** | 8 x 0.5 dB | 4 dB |
| **Feedback loop PUCs** | 2 x 0.5 dB | 1 dB |
| **Other PUCs** | 3 x 0.5 dB | 1.5 dB |
| **Total** | | 20.5-22.5 |

integrated implementation, in order to feed the two (four) nodes, as in this work each node was measured sequentially. Summing up, a two-node integrated ROSS receiver would exhibit losses fairly below 10 dB, which is a substantial improvement from the 22 dB of the present experiment.

## III. Results and Discussion

### *A. Evaluation of the concept in a 32 GBaud QAM-4/16 link*

In order to experimentally evaluate this idea, we implemented the setup of Fig. 1c. The modulated C-band signal is transmitted through single-mode fiber, amplified, filtered by the ROSS nodes, and then received by a simple photodetector. Different fiber transmission distances are considered. Due to practical equipment limitations, for this proof-of-concept experimental validation, the different ROSS nodes are implemented and measured sequentially by reconfiguring the photonic circuit. Digitization and equalization based on FFE or a simple Neural Network DSP follow the photodetection in order to demodulate the I and Q components of the signal. The optical spectrum of signal, after being filtered by two ROSS nodes, is depicted in Fig. 1d.

In Fig. 3, we experimentally showcase the principle of operation of the slicer and its intrinsic phase to intensity conversion properties which depend on the frequency detuning and the transfer function. The study has been carried out for a QAM-4 signal, transmitted through a 25-km standard single-mode fiber (SSMF) spool. Initially the two filters have no detuning ($\Delta f$=0 GHz) and the QPSK signal superimposed in the first quadrant degenerates to a three-state intensity signal as the intensities of symbols (-1, 1) and (1, -1) are identical, whilst symbol (1,1) and symbol (-1, -1) have the highest and lowest intensity respectively . When parameter $\Delta f$ is properly tuned, the full constellation, after a linear regression process, is revealed and becomes really an ideal QPSK signal for $\Delta f$=5.6 GHz. Fig. 3e) shows the optimization contour plot of $\Delta\varphi$ tuning for a two-node receiver, with respect to the BER for the same transmission system of 25 km reach. This plot is symmetrical, with the worst performance exhibited when $\Delta\varphi_1=\Delta\varphi_2$, which is analogous to a single node case. In a region where $\Delta\varphi 1 \approx 1$ rad and $\Delta\varphi 2 \approx 4$ rad, BER performance exhibits significant improvement relating with the fact that transmission impairments have been sufficiently mitigated (the noise is decolored) and the phase information has been properly imprinted into the amplitude domain. These phase shifts correspond to a frequency shift of around 3.5 GHz and 14 GHz respectively from the initial filter frequency, so the two filters are separated by a 10.5 GHz frequency shift. This 10.5 GHz divided by two is the final frequency detuning of each filter with respect to the signal carrier. Based on the wavelength of the carrier and the central frequencies of the two nodes, we conclude that for the proper operation of the ROSS accelerator the two nodes must exhibit approx. 5 GHz antisymmetric detuning around the carrier, as shown in Fig. 2c, which corresponds to 0.16xbandwidth of the signal. The optimality of the antisymmetric frequency detuning and the detuning value are in agreement with simulation results from previous works [15].

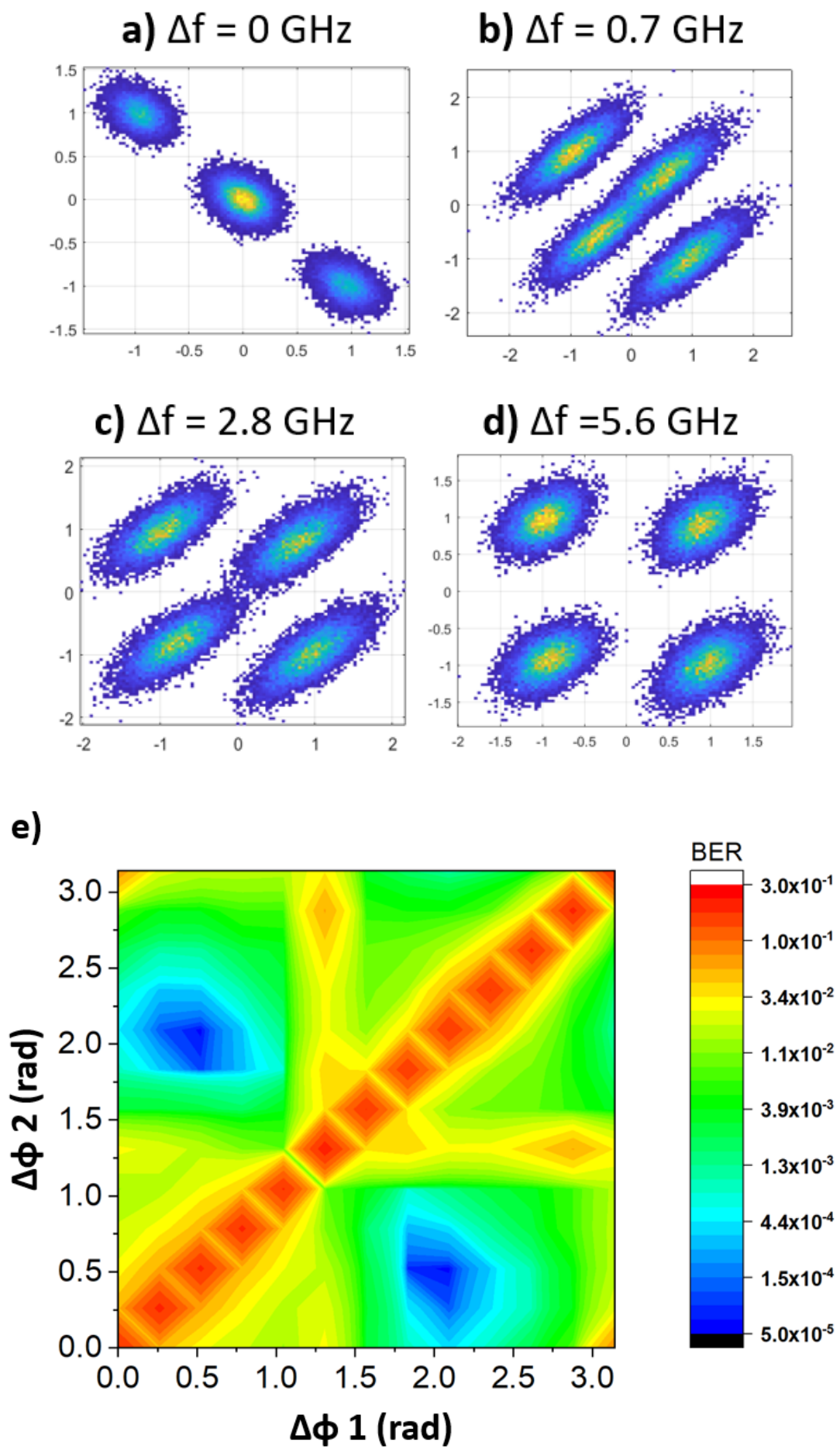

Figure 3. a) Minimum phase QAM-4 signal received by a simple photodetector (three amplitude levels are received instead of four). b-d) The same signal received by two antisymmetrically detuned, by Δf with respect to the carrier, recurrent filters and two photodetectors after linear equalization. e) The optimization process of the ROSS coherent receiver. The optimization is done in respect of the phase difference between the two arms of each MZDI, $\Delta\varphi$, for two nodes concurrently. Each rad of Δφ corresponds to about 3.6 GHz. There are regions in which the BER is reduced even by 4 orders of magnitude. This result refers to a 32 GBaud QAM-4 signal after 25 km of transmission.

After finding the best performing combination of nodes, we scan the critical modulator driving swing parameter. In Fig. 4a, we provide the BER results versus this driving voltage for each transmission distance examined in this work, in the case of QAM-4, with two ROSS nodes. We vary the driving swing up to the $V_\pi$ value of 3.5 V. One can notice the performance trade-off between the higher SNR that the larger modulation amplitudes offer and the higher nonlinearity as the driving swing grows. The optimal point is found to be around 1 V, for a small transmission distance (25 km), while it is transferred to 1.5 V and even 2 V for the cases of 50 km and 75 km due to the higher losses, which shifts the traded-off driving point towards improved SNR. BER below $10^{-5}$ is achieved in the back-to-back configuration, while even in the case of 75 km the performance remains in the order of $10^{-4}$. The same behavior can be noticed for the QAM-16 format, in Fig. 4b showing that the receiver is modulation format agnostic. In this case, we increase the ROSS nodes from two to four in order to enhance the frequency diversity required by the higher order modulation format. Hence, BER remains near $10^{-2}$ for up to 50 km, because the 75 km case corresponds to transmission losses of 15 dB which, when superimposed to the 22 dB losses of the Smartlight degrade severely the SNR.

In order to analyze the impact of limited SNR performance in our specific implementation, we employ digital averaging of up to 10 DSO captures. We study QAM-16 at 0.75 V, 1 V, 2 V and 3 V in order to gain understanding of how the scheme could perform for a dedicated, potentially lower-loss ROSS receiver. Fig. 4c showcases that the BER improvement for small modulation amplitudes approaches an order of magnitude (800%), with BER=$10^{-3}$ for an averaging of 10 captures at 750 mV. On the other hand, at the nonlinear cases of 2 V and 3 V the improvement does not surpass 40%. This result proves that, even at the back-to-back scenario, this experiment is noise-limited, mainly from the high loss of the Smartlight node, and the subsequent amplification required, and highlights the remaining potential for improved performance with custom designed, fabricated and packaged filters.

Another important aspect of the proposed receiver architecture for practical implementation refers to the tolerance to the laser frequency drifts. Frequency drifts in the input signals will affect the optimal positioning of the ROSS nodes' frequency detuning. In order to investigate the tolerance to this impairment, we start from the optimal combination of four nodes for the QAM-16 case and then concurrently shift them by the same $\Delta\varphi$, emulating laser's frequency drift. Fig. 5 illustrates the effect of this deviation on the BER. Due to the periodicity of the MZDI spectral response, we observe a similar periodic pattern in BER variations. However, BER remains almost constant for drifts below 1 GHz, while it deteriorates up to 50% for frequency deviations exceeding 3 GHz. This 1-3 GHz margin of BER tolerance to laser frequency drifts at the 32 Gbaud experiment can be expected to translate to 4-12 GHz tolerance at a 120 Gbaud scenario that could be considered for contemporary 800G transceivers. Based on this result, we expect that the proposed receiver could operate in a semi-cooled regime with wavelength stabilization of around 0.05 nm (6.25 GHz). Such loose stabilization is also considered even for IM/DD systems in high baudrates, in order to avoid the increasing CD at the edges of the O-band [20]. In any case, thermoelectric coolers (TEC) for coherent transmitters can easily achieve wavelength stability <2 pm (about +/- 0.02 $^0$C stability) with moderate power consumption (below 0.5 W even for high ambient temperatures) [21]. Another, more elegant approach for the system stabilization combines the essential monitoring of the recurrent filters phase shifters deviation, with the monitoring of the slow thermal drifts of the laser, thus avoiding the power-hungry usage of TEC controllers. For example, dithering schemes can offer reliable stabilization in entire meshes of phase shifters by dynamically tuning them, counterbalancing effectively environmental deviations [22]. In Fig. 5, we also provide results for a two-node receiver with moderate (x3) digital averaging, achieving BER=$2x10^{-2}$. Thus, we prove that the receiver can operate with the same number of receivers but with half the number of photodetectors as the coherent receiver under the proper SNR conditions.

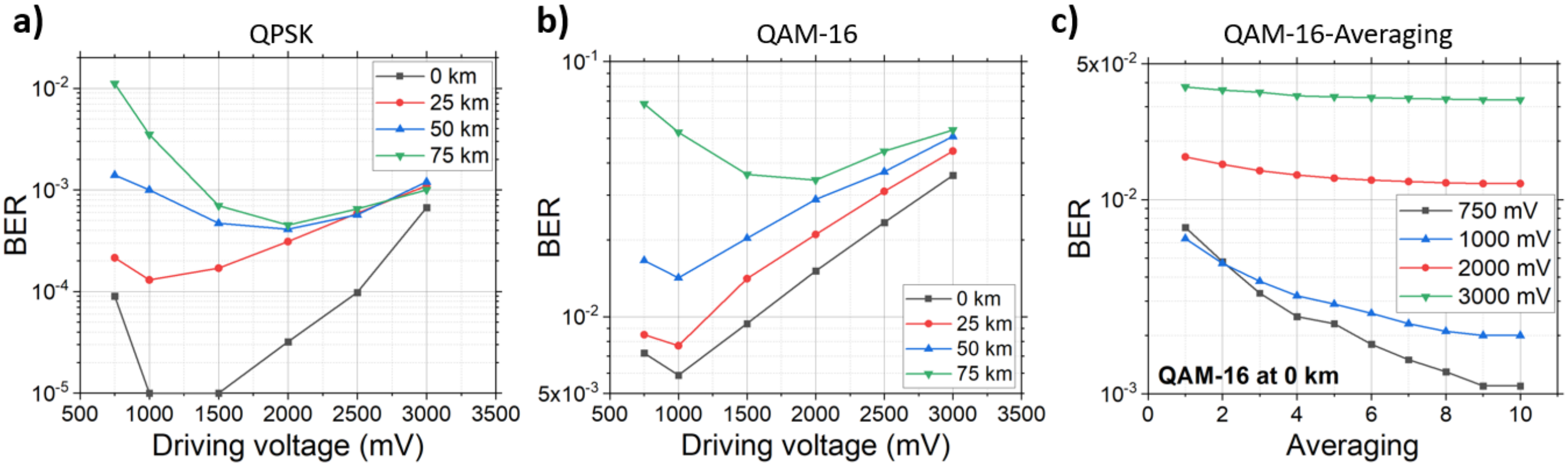

Figure 4. BER performance of the receiver. a) BER results versus the modulator driving voltage for QAM-4 with two nodes and b) for QAM-16 with four nodes, for different transmission distances. The best performance is noticed at around 1000 mV, which correspond to $0.3xV_{\pi}$. Lower voltage deteriorates SNR, while for values approaching Vπ, the constellation is affected by the modulator nonlinearity. c) BER versus the digital averaging for the QAM-16 case. For small driving voltages, with lower SNR, the digital averaging improves the performance by about an order of magnitude, proving that the experiment is mainly SNR limited.

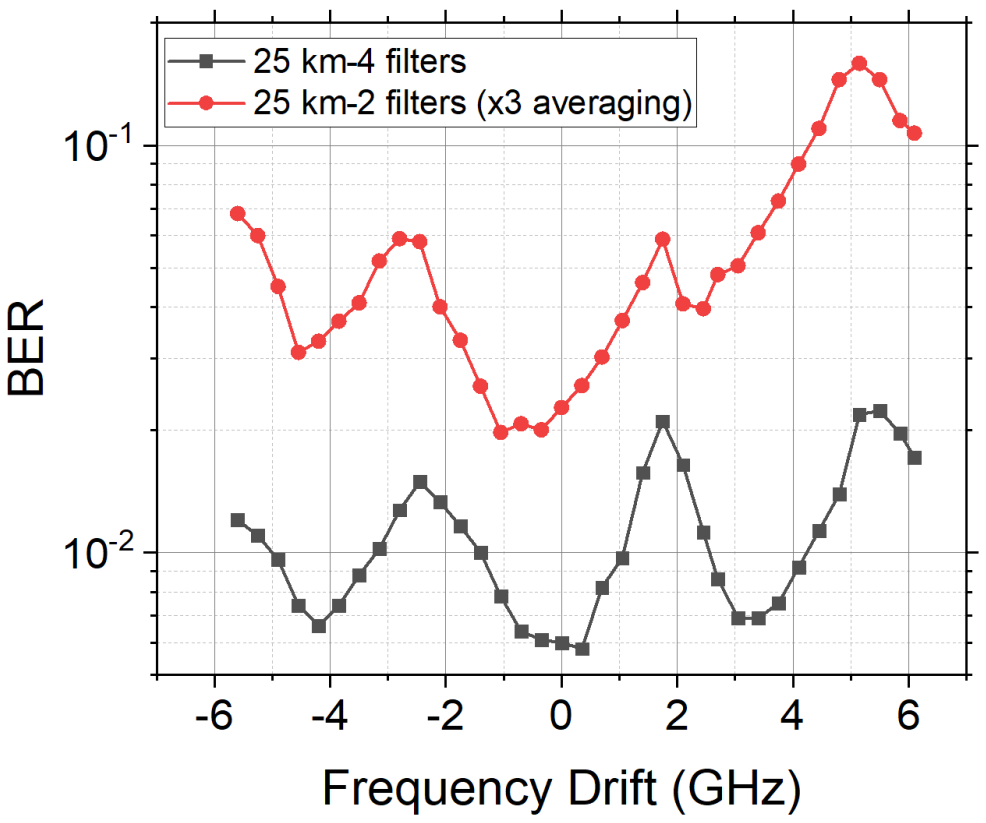

Figure 5. The tolerance of the transceiver to the frequency drift of the laser source. The four-node system performance for the QAM-16 case seems steady for frequency drifts over 1 GHz, while lies within a 50% deterioration region for over 3 GHz drifts. A receiver with two filters is also examined with the aid of moderate digital averaging, achieving also acceptable results.

### *B. Improvement through GCS*

The signal modulation with driving voltage near the $V_{\pi}$ value causes a nonlinear distribution of the symbols across the first IQ quadrant. In order to diminish the effects of this type of nonlinearity, and most importantly to properly tailor the nonlinear transformation of phase to amplitude in the recurrent filter (as shown in Fig. 6a), we use GCS. The proper position of each symbol can be the result of a gradient-free training of the modulation as has been proposed in previous works [23]. Here, in order to avoid complex methods, we indicatively perform blind density-based spatial clustering of applications with noise (DBSCAN) as a potential method for GCS. DBSCAN is designed to cluster data of arbitrary shapes in the presence of noise in spatial and non-spatial high-dimensional databases [24]. The key idea of DBSCAN is that for each object of a cluster, the neighborhood of a given radius $R$ has to contain at least a minimum number of objects *MinPts*, which means that the cardinality of the neighborhood has to exceed some threshold. If the $R$-neighborhoods of a point $P$ at least contain a minimal number of points, then this point is called a core point. The core point is defined as: $N_R(P) > MinPts$. Here $R$ and *MinPts* are the user's specified parameters, i.e. the radius of the $R$-neighborhood and minimum number of points in the neighborhood of a core point, respectively. If this condition is not satisfied, then this point is considered as a non-core point. In this work, we employ DBSCAN by searching the optimal set of $R$ and *MinPts* values in any separate case. Typical values for $R$ are lying below 0.4 due to the constellation distribution which ranges from -3 to 3 for QAM-16. *MinPts* are related to the total number of transmitted symbols, usually here *MinPts*<1000. For each received constellation, we search for the parameters that create 16 solid clusters and then find their centroids. The centroids contain little information about the noise distribution but constitute the simplest way to extract the inverse constellation. Finally, we apply the inverse constellation from the relation: $(New\ const.) = 2\ (Initial\ const.) - (Centroid\ const.)$. This new constellation is retransmitted in order to counterbalance the effects of the transmission system. More advanced GCS methods, taking into account the full noise distribution, are expected to further improve performance.

The basic concept of the followed method is presented in Fig. 6, where the initial signal characterized by nonlinearities is detected (Fig. 6a), clustered (Fig, 6b) and a constellation in which we pre-compensate for the modulator offset and the nonlinear phase to amplitude conversion (GCS in Fig. 6c) is retransmitted and received. In Fig. 6d, we present the received unoptimized and optimized constellations for the relatively linear case of the 1000 mV and the nonlinear case of 2500 mV driving swing. In the 1000 mV case, the transmission of the optimized constellation delivers a relatively linear result, while even for the nonlinear case the nonlinearity is partially mitigated. This is also highlighted in the BER results for 0 km (Fig. 6e), 25 km (Fig. 6f) and 50 km (Fig. 6g) transmission. The improvement is negligible for 750 mV, while increases subsequently and saturates after 1.5 V. For comparison, we include the results of a strong recurrent neural network (RNN) nonlinear equalization in the case of 0 km. More specifically, we employ bi-directional vanilla RNNs with 28 hidden units, exploiting the many-to-many training approach, with the input layer nodes equaling the number of FFE taps (21). The output of the RNN decodes simultaneously 15 symbols which secures good performance but also minimum complexity, according to the many-to-many approach [25]. Each neuron employs a tanh nonlinearity. The overall network complexity is calculated to 2856 parameters and 3998 multiplications per symbol. The RNN model is built, trained and evaluated in Keras with Tensorflow 2.3 GPU backend. In the Keras model, binary cross-entropy is chosen as a loss function and Adam as the optimizer. We considered a single input in the RNN model, that of the transmitted signal. We also consider 45000 symbols for training, 10000 for validation and 10000 for testing. The models and the training of our benchmark RNN equalizers are extensively discussed in [26]. The RNN equalizer and the DBSCAN optimization aim at mitigating the nonlinearity of the system and the RNN case provides better performance. Constellation shaping creates a good fit with the proposed receiver and can offer an appealing solution in SNR limited environments where there is need for increased driving voltages. Other GCS techniques or probabilistic constellation shaping [27] could also be investigated in future works.

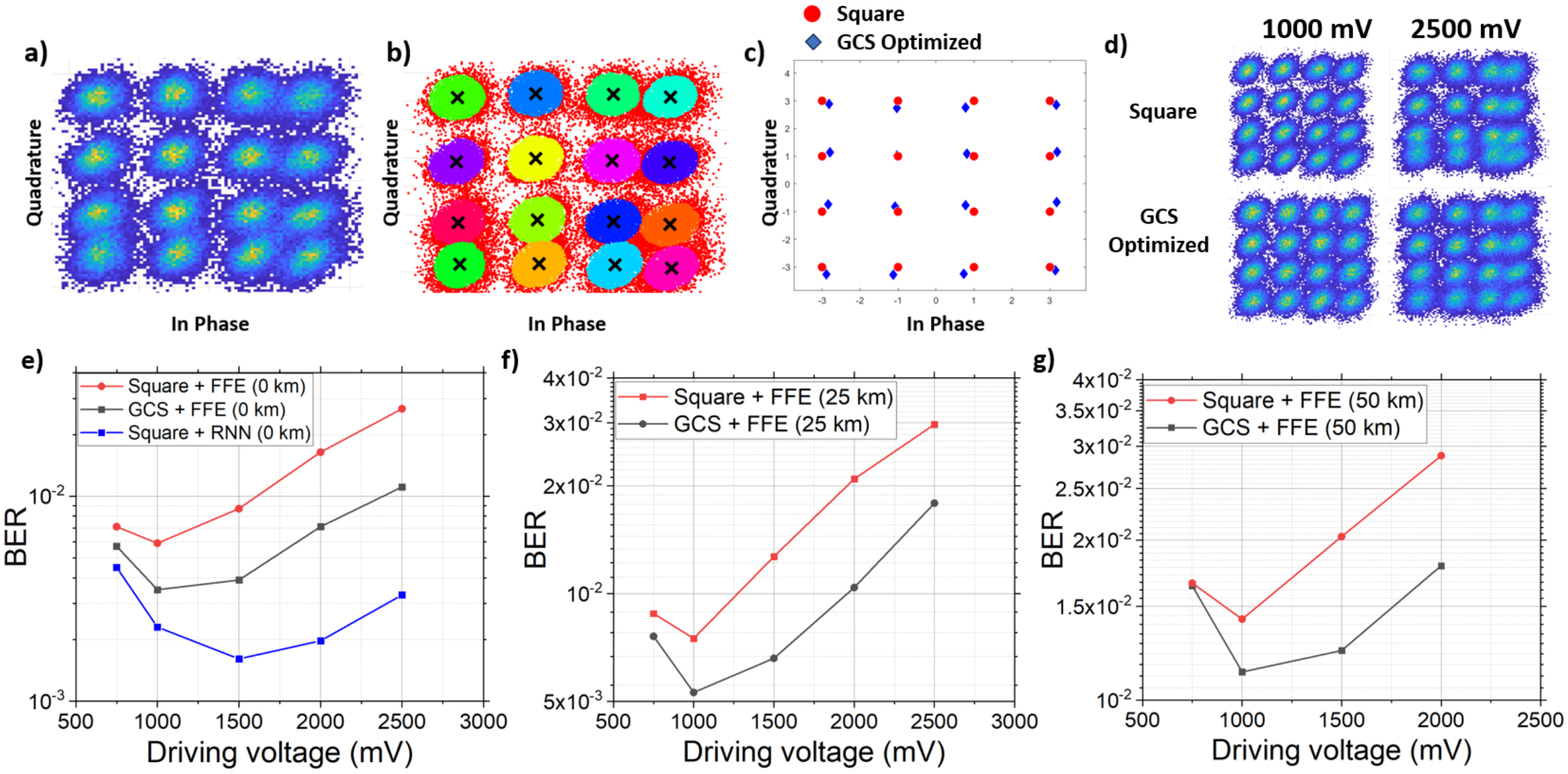

Figure 6. GCS optimization. The subfigures a)-c) illustrate the procedure of DBSCAN optimization of the transmitted constellation while d) showcases the improved received constellations for 1000 mV and 2500 mV. Figures e)-g) present the BER results for the different transmission distances. At the back-to-back case (0 km) a RNN equalizer is also used as a strong benchmark to DBSCAN optimization.

### *C. Power consumption analysis and discussion*

The proposed ROSS coherent transceiver aims to simplify both the transmitter and the receiver parts of the current coherent propositions for short-reach links, reducing significantly power consumption. Here, we compare three 1.6T LR transceiver variants in terms of power consumption, based on the next-generation 8x200G IM/DD scheme, the coherent 2x800G and an envisioned 4x400G ROSS self-coherent module. We exclude from this comparison the parts that are expected to be rather similar for all schemes, like FEC codes, transmitter DSP, photodetectors and trans-impedance amplifiers, as also DC converters, controllers etc. Table II, contains a part-by-part comparison of the power consumption, with the RF drivers, lasers, modulators and TEC at the Tx and the ADC/DSP of the Rx. Based on the results of this work, the ROSS transmitter, with driving voltages even lower from 1 V, is considered driver-free even for high $V_\pi$ silicon modulators, which are preferrable as the only mature technology at this stage. An indicative power budget is drawn in order to prove that the laser diodes power consumption does not substantially exceed that of the IM/DD scheme even with the addition of a dedicated silicon photonic chip for the ROSS receiver, due to the higher order modulation format which leads to half the number of lasers. The TEC stability is considered the same for all the three schemes (semi-cooled operation). This is because the IM/DD is expected to require a denser wavelength division multiplexing [28] as the chromatic dispersion grows at 100 Gbaud, while the coherent LR1 does not require the strict cooling conditions of the metro implementations. Previously, we showcased how the ROSS scheme can tolerate substantial frequency drifts in order to take advantage of the same TEC conditions with the other schemes. In the case of IM/DD and ROSS, we provide the power consumption with and without the

TABLE II

*The power consumption analysis of an 1.6T ROSS coherent transceiver in comparison with the two most prominent solutions of IM/DD and coherent technology.*

| **Architecture** | **8λ x 200 Gbps WDM IM/DD** | **2λ x 800 Gbps WDM Coherent-lite** | **4λ x 400 Gbps WDM ROSS coherent** |
|---|---|---|---|
| **Format** | **120 Gbaud PAM-4** | **120 Gbaud DP-16QAM** | **120 Gbaud 16QAM** |
| **RF Driver** | 225x8=1.8W | 225x8=1.8W | 0 W |
| **Laser** | 8×43 mW = 352 mW | 2x86 mW = 172 mW | 4x135 mW = 540 mW |
| **TEC** | 8x250 = 2 W or 0 W | 2x250 = 0.5W | 4x250 = 1 W or 0 W |
| **Modulator** | $V_{rms}$ = 1 V and P=280 mW | $V_{rms}$ = 2 V and P=940 mW | $V_{rms}$=0.35 V and P=200 mW |
| **ADC** | 8x360 mW = 2.88 W | 8x518 mW =4.144W or 8x360 mW = 2.88 W | 8x360 mW = 2.88 W |
| **Rx DSP** | 8x350mW=2.4 W | 2x(2300 mW) = 4.6 W | 4x400 mW = 1.6 W |
| **Total Power Consumption** | 9.71 W (7.71 W without TEC) | 12.154 W (10.89 W with baudrate sampling) | 6.22 W (5.22 without TEC) |

TEC contribution, as it could be avoided in various system implementations, substituted by phase shifter tracking mechanisms which consume marginal power. The power consumption of the modulator is calculated based on the $V_{rms}$ in the termination of the MZM and the phase shifters consumption [29]. For the ADCs, we consider the IM/DD and ROSS coherent scheme with the advantage of baudrate sampling (1 sps), while the standard coherent requires an oversampling of ≥1.2 sps. Lastly, we refer only to the signal extraction and equalization and omit the clock recovery, the FEC and any possible serialization/deserialization at the Rx DSP part. The power consumption analysis designates substantial gains for the ROSS scheme even in comparison with the low-consumption IM/DD scheme. This is attributed to the increasing requirements of IM/DD for TEC and DSP at higher baudrates. But, even considering uncooled operation for IM/DD, the ROSS coherent transceiver is favorable due to driverless operation and the relatively lighter DSP. The picture would not change even if small-gain drivers (in the order of 6 dB) are incorporated in the procedure. In comparison with the typical coherent transceivers, even in their most lightweight versions, the ROSS transceiver exhibits at least 4.5 W lower power consumption margin, corresponding to more than 40% improvement, if we count only the contributors that are not assumed similar.

TABLE III

*Main power contributors in pluggable modules*

| **Power contributors** | **Assumed similar contributors** | **Same for all contributors** |
|---|---|---|
| RF drivers | FEC | Trans-impedance Amplifiers |
| Lasers | Transmitter DSP | Photodetectors |
| TEC | DACs | Serializer/Deserializer |
| Modulators | | DC conversions |
| ADCs | | Controllers |
| Receiver DSP | | |

### *D. Electrical spectral efficiency*

The electrical spectral efficiency (ESE) is defined as the achievable information rate, measured in bits/second, per frequency unit (Hz) of analog bandwidth [6]. In practice, ESE can be calculated as the ratio between the bitrate and the analog bandwidth of the receiver side components, like the PD and the ADC. Typical IM/DD systems with PAM-4 modulation and receiver bandwidth of ~0.4 times the baudrate, achieve ESE of ~5 bits/s/Hz. On the other hand, coherent schemes with QAM-16 and receiver bandwidth ~0.5 times the baudrate offer ~8 bits/s/Hz per polarization. Most self-coherent proposals based on QAM-16 offer ESE between 4 bits/s/Hz and 8 bits/s/Hz per polarization [6]. Here, we showcase the potential ESE improvement of the proposed transceiver by exploring the limits of analog bandwidth reduction in the receiver side. For this investigation, we further process the experimental results by reducing the bandwidth of the digital filter ($8^{th}$ order Butterworth) that we place after the digital signal oscilloscope, in order to emulate a bandwidth limited receiver. The results of Fig. 7, illustrate the performance degradation as the Rx bandwidth reduces. Starting from BER of $8.5x10^{-3}$ for the four-node, QAM-16 case at the 25 km reach, the performance remains below $10^{-2}$ for as low as 15 GHz of Rx bandwidth. The degradation becomes faster below the Nyquist bandwidth, but remains below $1.3x10^{-2}$ even for 11 GHz bandwidth, rising to BER=$1.9x10^{-2}$ at 10 GHz. Assuming Rx bandwidth of 0.35xbaudrate and QAM-16, the ESE is 11.4 bits/s/Hz for single polarization, increased by 42% with respect to the 8 bits/s/Hz of a coherent-lite scheme operating at baudrate sampling with analog bandwidth of 0.5xbaudrate. The ESE-performance trade-off can be leveraged in order to achieve ESE >13 bits/s/Hz/pol with BER<$2.5x10^{-2}$.

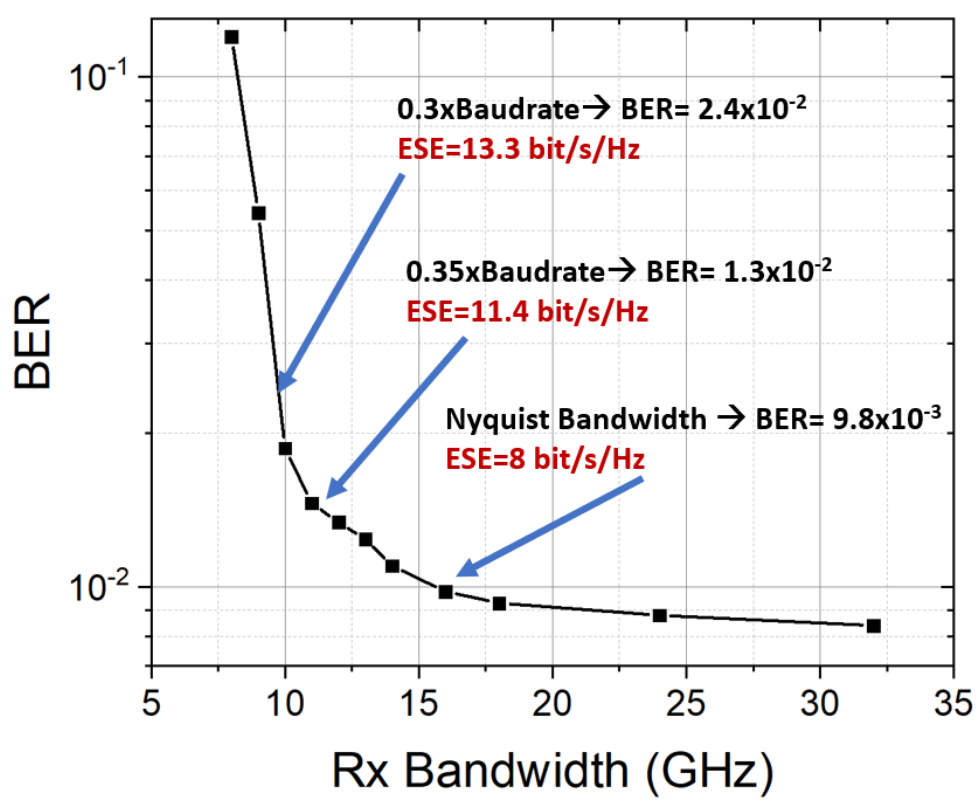

Figure 7. The Rx analog bandwidth and the ESE of the transceiver. The performance remains relatively stable for as low as 11 GHz, which correspond to 0.35 times the baudrate, providing ESE of 11.4 bits/s/Hz in the case of QAM-16.

### *E. Discussion*

In this work, we examine a single node configuration, mainly dictated from the capabilities of the Smartlight platform. However, parameters of the ROSS node such as the feedback loss, the feedback delay and the bandwidth play significant role

TABLE IV

*Indicative power budget of the three schemes under comparison*

| IM/DD | Coherent | ROSS |
|---|---|---|
| Minimum Link budget:<br>SiP chip insertion loss (IL) = 6.5 dB<br>(2 dB coupling and 4.5 dB MZM IL)<br>Modulation loss = 1 dB<br>Optical Mux/Demux IL = 4 dB<br>10 km fiber = 3.5 dB | Minimum Link budget:<br>SiP chip IL = 12 dB<br>(2 dB coupling, 6 dB DP IQM, 2 dB for the optical hybrid)<br>Modulation loss = 8 dB<br>LO = 3 dB (Splitting ratio 1:1)<br>Optical Mux/Demux IL = 4 dB<br>10 km fiber = 3.5 dB | Minimum Link budget:<br>SiP chip IL = 7 dB<br>(2 dB coupling, 5 dB IQM)<br>ROSS IL = 6.5 dB<br>Modulation loss = 2 dB<br>Optical Mux/Demux IL = 4 dB<br>10 km fiber = 3.5 dB |
| Total budget=15 dB<br>**Required** Optical output power=5 dBm<br>Laser Diode=43 $mW$ | Total budget=30.5 dB<br>Required Optical output power=10.5 dBm<br>Laser Diode=86 $mW$ | Total budget=23 dB<br>Required Optical output power=13 dBm<br>Laser Diode=135 $mW$ |

in the nonlinear phase-to-amplitude transformation of the incoming signal as previous works have demonstrated [15], [16]. In order to keep the MZDI losses within reasonable limits, we chose the minimalistic architecture which employed two and six PUCs as the two arms offering an FSR of 22 GHz, while other implementations could offer double the FSR at the expense of more complicated structures employing more PUCs. We employed the minimum possible feedback loop for our setup, that of the two PUCs, in order to restrain the feedback losses. Furthermore, in section II.B, we provided specific numbers about the losses in each stage of the Smartlight node implementation, illustrating how we can reduce the overall 22 dB to less than 10 dB with an application specific integrated filter node.

Another important topic is the thermal stability of the phase shifters in the unbalanced MZDI filters. Based on the results of section III.A and especially the Fig. 5, we proved that a tolerance for the frequency drifts in the order of few GHz would not practically affect the performance of the receiver. Today, there are many propositions for automatic phase error correction circuitries that can render the filter stable and immune to reasonable temperature variations, well below the GHz regime. For, example, in [30], a method consuming only 2.5 mW per channel is proposed for a complex 16-channel cascaded Mach-Zehnder wavelength filter. Another way for keeping the filters stable, while also monitoring the slow thermal drifts of the laser, is the incorporation of dithering schemes, as discussed earlier. Finally, even simpler tracking mechanisms, such as the received optical power tracking with a kHz rate (faster than the slow thermal drifts), would achieve real-time recalibration of the phase shifters, when the optimal frequency diverges more than ~1 GHz, with marginal computational effort.

## IV. Conclusion

This work provides evidence about the demodulation of M-QAM coherent formats with direct detection and IM/DD DSP enabled by the frequency processing offered by ROSS neuromorphic accelerators. The successful linearization of the 32 Gbaud, 50 km channel in the C-band can be considered dispersion equivalent with 128 Gbaud transmission of over 20 km in the O-band. The proposed transceiver can operate driver-free even with silicon photonic modulators and can take advantage of cheap and mature IM/DD components like medium linewidth distributed feedback lasers, low bandwidth photodetectors, trans-impedance amplifiers and ADCs, and drastically simplifies the DSP of coherent and self-coherent proposals. Moreover, it is compatible with polarization multiplexing which is regaining attention for IM/DD systems through optical endless polarization tracking [31], [32]. The power consumption gain per 1.6 Tb/s module could exceed 1.5 W with respect to IM/DD solutions and 4-6 W from their coherent counterparts.

## Appendix

In this appendix we provide the analysis regarding the power consumption comparison that is presented in this work between the proposed ROSS coherent transceiver and two of the most prominent solutions for 800G and 1.6T short-reach links [33]. The comparison is made upon the assumption of a 1.6T module based in three schemes. Firstly, in the 8x200 Gb/s/λ IM/DD, which is the straightforward sequel of the current state-of-the-art 4x200G. Secondly, in the 2x800 Gb/s/λ coherent based on the integration of two 800G transceivers in the same pluggable and finally to a 4 λ ROSS coherent module based on a two-filter receiver per λ. All three propositions are based on 105-125 Gbaud symbol rates and thus share the same class of optics. The comparison is constrained to only six power contributors of the module, while the rest of them are considered the same for all configurations, as shown in Table III. The assumptions that are considered for each one of them, leading to the final power consumption calculation are discussed below.

### *RF drivers*

For the IM/DD case, we assume a typical $V\pi$ of 4 V for the SiP Mach-Zehnder modulator, which would require about 3 V driving voltage for being in the linear region and a subsequent electrical amplification of about 10 dB, based on about 1 V voltage from the DAC. For the coherent modulator, the driving voltage is assumed the double of that of the IM/DD, so we calculate a required electrical gain of 16 dB. Finally for the proposed ROSS receiver, with Vpp=Vπ/4 or about 1 V, we assume driver-less operation. For our calculation we use the 4-channel linear driver with 48 GHz bandwidth and 13-22.5 dB tunable gain [34], consuming 225 mW per channel, which is lower than other commercial specifications [35]. So, the 8 channels of IM/DD or 2 channels of the polarization-multiplexed coherent correspond to 225x8=1.8 W.

### *Lasers*

Laser power consumption is estimated as $P_{LD}$=$V_{LD}$ $I_{LD,driv}$ and $I_{LD,driv}$=$P_{LD,opt}/\gamma_{LD}$ +$I_{LD,th}$. Here, $V_{LD}$ is the diode bias voltage, $I_{LD,driv}$ is the diode current, $P_{LD,opt}$ is the optical power of the laser, $\gamma_{LD}$ is the slope efficiency, and $I_{LD,th}$ is the threshold current. Although these parameters differ among vendors, a typical set of values is $V_{LD,f} = 1.6\,V$, $I_{LD,th}$=15 $mA$ and $\gamma_{LD}$=0.29 $mW/mA$ [36]. For approximating the $P_{LD,opt}$, we draw a rough power budget of each one of the three schemes, assuming Rx sensitivity of -10 dBm for IM/DD, as IM/DD systems require substantially increased received optical power for the same BER and -20 dBm for the coherent receiver at 120 Gbaud, based on [3]. Also, we assume typical modulation losses (see Table IV) as recently reported in literature [3]. It must be noted that coherent modulation induces high modulation losses, at least 8 dB more than the IM/DD at their full swing. Also, the laser power is split to the transmitter and the local oscillator (LO) at a ratio 1:1. Typically, this ratio is 2:1, but for the low output power case of this short reach application, less power is directed to the transmission. Lastly, ROSS transmitter will impose about 2 dB of modulation losses due to the assumed $V_\pi/4$ swing [37] and the ROSS chip will impose less than 6.5 dB of insertion losses [38]. The power budget is shown in Table IV.

*TEC*

The TEC power consumption is proportional to the number of lasers. For each laser we assume 250 mW for semi-cooled operation and simple wavelength locking with acceptable drifts up to 12.5 GHz [8], [21]. However, IM/DD systems may operate without TEC even at 100 Gbaud. For this reason, we calculate the total power consumption with and without the TEC contribution for the IM/DD scenario. Assuming IM/DD with TEC, we have 8 lasers and 2 W power consumption. Coherent requires two lasers, or 500 mW and the ROSS transceiver with 4 lasers requires 1 W. We also provide the power consumption of ROSS with and without TEC, as many mechanisms could supersede it, as discussed in the discussion part. Also, in the case of the frequency drift of the filters, we do not include the filter stability mechanisms in this power consumption comparison, as they consume a marginal amount of power in the order of mW, either in the dithering schemes or in any alternative solution.

*Modulator*

Here, we assume that a series of push pull MZM / IQM is used, and the driving swing depends on the architecture and modulator platform. For IM/DD, the power consumption in the modulator is given by $P_{modulator} = P_{MZM} + P_{\pi,TPS}/2$ , where $P_{MZM}$ is the power consumed in the RF termination and equals $P_{MZM} = V_{rms}^2/R$ with $V_{rms}$=0.35 $V_{pp}$=1 V and with R typically equal to 50 Ω [29]. $P_{\pi,TPS}$ is the power consumed by the thermal phase shifter to induce a phase-shift of $\pi$ as bias, with $P_{\pi/2,TPS}$ referring to a $P_{\pi/2,TPS}$= $P_{\pi,TPS}/2$ power . For coherent transmission, there are 4 MZMs and 2 additional phase shifters in the DP IQM. Thus, it consumes $P_{total} = 4\times P_{MZM}+4\times P_{\pi,TPS}+$ 2x $P_{\pi/2,TPS}$ with assumed $V_{pp}$= 1.5$V_\pi$ = 6 V. The single polarization coherent transmitter for ROSS scheme utilizes 2 MZMs and consumes $P_{total}$ = 2×$P_{MZM}$+3×$P_{\pi/2,TPS}$ with $V_{pp}$=$V_\pi$/4. We also consider $P_{\pi,TPS}/2$ = 10 mW for biasing IM/DD and ROSS at π/2 while $P_{\pi,TPS}$ = 20 mW [39] for biasing the coherent IQ modulators at the null point. Thus, the IM/DD modulator with $V_{rms}$=1 V requires $P$=30 mW per modulator or 240 mW for the 8 channels. The coherent case with $V_{rms}$=2 V requires P=420 mW per DP-IQ modulator or 840 mW for the two channels. Finally, the ROSS module, with $V_{rms}$=0.35 V, P=35 mW per IQ modulator or 140 mW for the 4 channels.

*ADC*

The ADC is one of the main sources of power consumption with its energy dissipation being proportional to the sampling rate, or even super-linear for very high data rates. Here, we assume a linear relationship for simplicity. Also, power consumption is related to the resolution with many figures of merit describing how the power doubles, triples or even being multiplied by a factor of 8 for every additional bit. Baud-rate sampling, synchronized to the symbol timing, improves the samples' peak-to-average power ratio and allows for reduction in ADC effective resolution by ½ bit, which correspond to at least 33% power savings [40]. For this comparison, we assume 300 mW for a 100 Gs/s ADC based on recent literature [41]. Moreover, we assume that IM/DD and ROSS transceivers take advantage of baudrate sampling with 120 Gs/s sampling rate, translated to 360 mW. For the coherent transceiver case, we assume an ambitious oversampling factor of 1.2 sps [42] and a moderate factor of 1.2 for the increased resolution, leading to 518.4 mW. However, we also present the total power consumption of the coherent module with the assumption of baud-rate sampling. All three schemes require 8 ADCs.

*Rx DSP*

Finally, for the receiver-side DSP, we include only the demodulation and equalizations algorithms in this comparison. For the IM/DD case, we assume a 21-tap feed-forward equalizer (FFE) along with a partial response equalizer and a maximum likelihood sequence estimator (MLSE) with 1-tap. This combination constitutes the simpler MLSE based equalizer, which is considered as the bare minimum for the 100 Gbaud IM/DD systems [43]. The complexity of the one-tap MLSE is equal to 1x$4^2$ real multiplications [44], the one of the FFE complexity equals to 1 real multiplication and an addition per tap, while we ignore here the partial response equalizer complexity. We can estimate the consumption of a digital algorithm, by the number of multiplications and the consideration of 0.3 mW for a multiplication at 8-bit resolution and 4 GHz rate at 0.7 V and 7 nm node [45]. With 9 mW per multiplication at 120 Gbaud, the 37 multiplications (at least) of the FFE+MLSE equalization consume about 330 mW per channel. For the ROSS receiver, we assume a 2x2 real valued FFE with 11 taps for 10 km reach, or 44 multiplications which correspond to 378 mW per channel. In the coherent case, we include only the complex 2x2 multi-input-multi-output equalizer with 11 taps [46] for the chromatic dispersion and the polarization mode dispersion compensation as also for the polarization demultiplexing. We assume no bulk dispersion compensation, because its frequency domain implementation exhibits higher complexity for number of taps above 13 [46]. The complexity of this equalizer is 4 complex multiplications per tap [47], which correspond to 16 real multiplications (plus two additions) , leading to over 1.5 W total consumption. We also have to include the carrier phase estimation and the frequency offset estimation that the coherent receiver requires, which are assumed to consume at least 800 mW [48].

## References

[1] D. Tauber *et al.*, "Role of Coherent Systems in the Next DCI Generation," *J. Lightwave Technol.*, vol. 41, no. 4, pp. 1139–1151, Feb. 2023, doi: 10.1109/JLT.2023.3235820.

[2] CFP MSA and D. Lewis, "CFP8 Hardware Specification," Revision 1.0, 2017. [Online]. Available: https://ascentoptics.com/storage/uploads/files/10000/20230217/cde62d886036e27b1178cad8e382700b.pdf

[3] X. Zhou, R. Urata, and H. Liu, "Beyond 1 Tb/s Intra-Data Center Interconnect Technology: IM-DD or Coherent?," *Journal of Lightwave Technology*, vol. 38, no. 2, pp. 475–484, Jan. 2020, doi: 10.1109/JLT.2019.2956779.

[4] T. Gui, Q. Xiang, and L. Xiang, "Feasibility Study on Baud-Rate Sampling and Equalization (BRSE) for 800G-LR1," San Antonio, Texas, IEEE 802.3dj May Interim meeting, Aug. 2023. [Online]. Available: https://www.ieee802.org/3/dj/public/23_05/gui_3dj_01a_2305.pdf

[5] D. Che, A. Li, X. Chen, Q. Hu, Y. Wang, and W. Shieh, "Stokes vector direct detection for short-reach optical communication," *Optics Letters*, vol. 39, no. 11, p. 3110, Jun. 2014, doi: 10.1364/OL.39.003110.

[6] D. Che, C. Sun, and W. Shieh, "Optical Field Recovery in Stokes Space," *J. Lightwave Technol.*, vol. 37, no. 2, pp. 451–460, Jan. 2019, doi: 10.1109/JLT.2018.2879181.

[7] A. Mecozzi, C. Antonelli, and M. Shtaif, "Kramers–Kronig coherent receiver," *Optica*, vol. 3, no. 11, p. 1220, Nov. 2016, doi: 10.1364/OPTICA.3.001220.

[8] F. Chang and Rang-Chen, "Relative Cost Analysis on IM-DD vs Coherent for 800G-LR, IEEE P802.3df," 2022. [Online]. Available: https://www.ieee802.org/3/df/public/22_11/chang_3df_01_2211.pdf

[9] T. Bo and H. Kim, "Toward Practical Kramers-Kronig Receiver: Resampling, Performance, and Implementation," *Journal of Lightwave Technology*, vol. 37, no. 2, pp. 461–469, Jan. 2019, doi: 10.1109/JLT.2018.2869733.

[10] D. Che, C. Sun, and W. Shieh, "Maximizing the spectral efficiency of Stokes vector receiver with optical field recovery," *Opt. Express*, vol. 26, no. 22, p. 28976, Oct. 2018, doi: 10.1364/OE.26.028976.

[11] R. Zhang, K. Kuzmin, Y.-W. Chen, and W. I. Way, "800G/$\lambda $ Self-Homodyne Coherent Links With Simplified DSP for Next-Generation Intra-Data Centers," *J. Lightwave Technol.*, vol. 41, no. 4, pp. 1216–1222, Feb. 2023, doi: 10.1109/JLT.2022.3218764.

[12] H. Chen, N. K. Fontaine, J. M. Gené, R. Ryf, D. T. Neilson, and G. Raybon, "Full-Field, Carrier-Less, Polarization-Diversity, Direct Detection Receiver based on Phase Retrieval," Mar. 2019, [Online]. Available: http://arxiv.org/abs/1903.02424

[13] B. Chen *et al.*, "Joint Optimization of Phase Retrieval and Forward Error Correcting for Direct Detection Receiver," in *2020 European Conference on Optical Communications (ECOC)*, Brussels, Belgium: IEEE, Dec. 2020, pp. 1–4. doi: 10.1109/ECOC48923.2020.9333228.

[14] X. Li, M. OaSullivan, Z. Xing, M. E. Mousa-Pasandi, and D. V. Plant, "Asymmetric Self-Coherent Detection Based on Mach-Zehnder Interferometers," *J. Lightwave Technol.*, vol. 40, no. 7, pp. 2023–2032, Apr. 2022, doi: 10.1109/JLT.2021.3135000.

[15] K. Sozos, A. Bogris, P. Bienstman, G. Sarantoglou, S. Deligiannidis, and C. Mesaritakis, "High-speed photonic neuromorphic computing using recurrent optical spectrum slicing neural networks," *Communications Engineering*, vol. 1, no. 1, p. 24, Oct. 2022, doi: 10.1038/s44172-022-00024-5.

[16] K. Sozos, S. Deligiannidis, C. Mesaritakis, and A. Bogris, "Self-Coherent Receiver Based on a Recurrent Optical Spectrum Slicing Neuromorphic Accelerator," *J. Lightwave Technol.*, pp. 1–9, 2023, doi: 10.1109/JLT.2023.3235278.

[17] I. Alimi *et al.*, "A Review of Self-Coherent Optical Transceivers: Fundamental Issues, Recent Advances, and Research Directions," *Applied Sciences*, vol. 11, no. 16, p. 7554, Aug. 2021, doi: 10.3390/app11167554.

[18] G. Tzimpragos, C. Kachris, I. B. Djordjevic, M. Cvijetic, D. Soudris, and I. Tomkos, "A Survey on FEC Codes for 100 G and Beyond Optical Networks," *IEEE Commun. Surv. Tutorials*, vol. 18, no. 1, pp. 209–221, 2016, doi: 10.1109/COMST.2014.2361754.

[19] D. Pérez-López, A. López, P. DasMahapatra, and J. Capmany, "Multipurpose self-configuration of programmable photonic circuits," *Nat Commun*, vol. 11, no. 1, p. 6359, Dec. 2020, doi: 10.1038/s41467-020-19608-w.

[20] D. Che and X. Chen, "Modulation Format and Digital Signal Processing for IM-DD Optics at Post-200G Era," *J. Lightwave Technol.*, vol. 42, no. 2, pp. 588–605, Jan. 2024, doi: 10.1109/JLT.2023.3311716.

[21] Z. Chen *et al.*, "Nano-ITLA Based on Thermo-Optically Tuned Multi-Channel Interference Widely Tunable Laser," *J. Lightwave Technol.*, vol. 41, no. 16, pp. 5405–5411, Aug. 2023, doi: 10.1109/JLT.2023.3261897.

[22] W. Zhang *et al.*, "Silicon microring synapses enable photonic deep learning beyond 9-bit precision," *Optica*, vol. 9, no. 5, p. 579, May 2022, doi: 10.1364/OPTICA.446100.

[23] O. Jovanovic, M. P. Yankov, F. Da Ros, and D. Zibar, "Gradient-Free Training of Autoencoders for Non-Differentiable Communication Channels," *J. Lightwave Technol.*, vol. 39, no. 20, pp. 6381–6391, Oct. 2021, doi: 10.1109/JLT.2021.3103339.

[24] H. V. Singh, A. Girdhar, and S. Dahiya, "A Literature survey based on DBSCAN algorithms," in *2022 6th International Conference on Intelligent Computing and Control Systems (ICICCS)*, Madurai, India: IEEE, May 2022, pp. 751–758. doi: 10.1109/ICICCS53718.2022.9788440.

[25] Y. Osadchuk *et al.*, "Low-Complexity Samples Versus Symbols-Based Neural Network Receiver for Channel Equalization," *J. Lightwave Technol.*, vol. 42, no. 15, pp. 5167–5174, Aug. 2024, doi: 10.1109/JLT.2024.3390227.

[26] S. Deligiannidis, C. Mesaritakis, and A. Bogris, "Performance and Complexity Analysis of Bi-Directional Recurrent Neural Network Models Versus Volterra Nonlinear Equalizers in Digital Coherent Systems," *Journal of Lightwave Technology*, vol. 39, no. 18, pp. 5791–5798, Sep. 2021, doi: 10.1109/JLT.2021.3092415.

[27] J. Cho and P. J. Winzer, "Probabilistic Constellation Shaping for Optical Fiber Communications," *J. Lightwave Technol.*, vol. 37, no. 6, pp. 1590–1607, Mar. 2019, doi: 10.1109/JLT.2019.2898855.

[28] R. Chen and F. Chang, "Feasibility of 800G LR4 and 800G ER8 with PAM4 IMDD, IEEE P802.3df," 2022. [Online]. Available: https://www.ieee802.org/3/df/public/22_03/yu_3df_01a_220329.pdf

[29] E. Berikaa *et al.*, "Next-Generation O-Band Coherent Transmission for 1.6 Tbps 10 km Intra-Datacenter Interconnects," *J. Lightwave Technol.*, vol. 42, no. 3, pp. 1126–1135, Feb. 2024, doi: 10.1109/JLT.2023.3307504.

[30] L. Han, B. P.-P. Kuo, A. Pejkic, N. Alic, and S. Radic, "50GHz Silicon Cascaded Mach-Zehnder Wavelength Filter and Automatic Phase Error Correction," in *Optical Fiber Communication Conference (OFC) 2019*, San Diego, California: OSA, 2019, p. W3B.3. doi: 10.1364/OFC.2019.W3B.3.

[31] D. Che and X. Chen, "Single-wavelength 1.2-Tb/s IM-DD transmission by polarization-multiplexing three 160-GBd PAM-8 signals," in *49th European Conference on Optical Communications (ECOC 2023)*, Hybrid Conference, Glasgow, UK: Institution of Engineering and Technology, 2024, pp. 1690–1693. doi: 10.1049/icp.2023.2668.

[32] C. R. Doerr, "Dual-polarization direct-detection for data-center optical communications," in *Next-Generation Optical Communication: Components, Sub-Systems, and Systems XII*, G. Li, K. Nakajima, and A. K. Srivastava, Eds., San Francisco, United States: SPIE, Mar. 2023, p. 47. doi: 10.1117/12.2653256.

[33] H. Isono, "Latest standardization trend and future prospects for 800G/1.6T optical transceivers," in *Next-Generation Optical Communication: Components, Sub-Systems, and Systems XII*, G. Li, K. Nakajima, and A. K. Srivastava, Eds., San Francisco, United States: SPIE, Mar. 2023, p. 14. doi: 10.1117/12.2648120.

[34] T. Jyo, M. Nagatani, J. Ozaki, M. Ishikawa, and H. Nosaka, "12.3 A 48GHz BW 225mW/ch Linear Driver IC with Stacked Current-Reuse Architecture in 65nm CMOS for Beyond-400Gb/s Coherent Optical Transmitters," in *2020 IEEE International Solid- State Circuits Conference - (ISSCC)*, San Francisco, CA, USA: IEEE, Feb. 2020, pp. 212–214. doi: 10.1109/ISSCC19947.2020.9063027.

[35] "MACOM", Accessed: Sep. 13, 2024. [Online]. Available: https://www.macom.com/files/live/sites/macom/files/Brochures/MACOM_Opto-brochure_web.pdf

[36] "Aerodiode model 3 (O-band) DFB laser", Accessed: Sep. 13, 2024. [Online]. Available: https://www.aerodiode.com/product/1310-nm laser-diode

[37] H. Zhang, "Power Efficient Coherent Detection for Short-Reach System," in *Optical Fiber Communication Conference (OFC) 2023*, San Diego California: Optica Publishing Group, 2023, p. M1E.1. doi: 10.1364/OFC.2023.M1E.1.

[38] I. Teofilovic *et al.*, "Integrated Recurrent Optical Spectral Slicer for Equalization of 100-km C-band IM/DD Transmission," in *2025 European Conference on Optical Communications (ECOC)*, Copenhagen, Denmark: IEEE, Sep. 2025, pp. 1–4. doi: 10.1109/ECOC66593.2025.11263068.

[39] S. Liu *et al.*, "Thermo-optic phase shifters based on silicon-on-insulator platform: state-of-the-art and a review," *Front. Optoelectron.*, vol. 15, no. 1, p. 9, Dec. 2022, doi: 10.1007/s12200-022-00012-9.

[40] O. Vidal and T. C. Carusone, "Benefits of a Coherent Solution Tailored for 800G-LR1," May 2023. Accessed: Sep. 13, 2024. [Online]. Available: https://www.ieee802.org/3/dj/public/23_05/carusone_3dj_01_2305.pdf

[41] R. L. Nguyen *et al.*, "8.6 A Highly Reconfigurable 40-97GS/s DAC and ADC with 40GHz AFE Bandwidth and Sub-35fJ/conv-step for 400Gb/s Coherent Optical Applications in 7nm FinFET," in *Digest of Technical Papers - IEEE International Solid-State Circuits Conference*, Institute of Electrical and Electronics Engineers Inc., Feb. 2021, pp. 136–138. doi: 10.1109/ISSCC42613.2021.9365746.

[42] W.-J. Jiang, K. G. Kuzmin, and W. I. Way, "Effect of Low Over-Sampling Rate on a 64Gbaud/DP-16QAM 100-km Optical Link," *IEEE Photonics Technology Letters*, vol. 30, no. 19, pp. 1671–1674, Oct. 2018, doi: 10.1109/LPT.2018.2864639.

[43] R. Nagarajan, I. Lyubomirsky, J. 13, and C. Webcast, "Inphi Moves Big Data Faster Next-Gen Data Center Interconnects: The Race to 800G."

[44] Y. Yu, Y. Che, T. Bo, D. Kim, and H. Kim, "Reduced-state MLSE for an IM/DD system using PAM modulation," *Optics Express*, vol. 28, no. 26, p. 38505, Dec. 2020, doi: 10.1364/oe.410674.

[45] R. K. Pandey and S. K. Pandey, "Analyzing the Performance of 7nm FinFET Based Logic Circuit for the Signal Processing in Neural Network," in *2020 IEEE Recent Advances in Intelligent Computational Systems, RAICS 2020*, Institute of Electrical and Electronics Engineers Inc., Dec. 2020, pp. 136–140. doi: 10.1109/RAICS51191.2020.9332500.

[46] C. Laperle and M. Osullivan, "Advances in high-speed DACs, ADCs, and DSP for optical coherent transceivers," *Journal of Lightwave Technology*, vol. 32, no. 4, pp. 629–643, Feb. 2014, doi: 10.1109/JLT.2013.2284134.

[47] X. Zhang *et al.*, "Real time low-complexity adaptive channel equalization for coherent optical transmission systems," *Optics Express*, vol. 28, no. 4, p. 5058, Feb. 2020, doi: 10.1364/oe.385370.

[48] E. Borjeson, C. Fougstedt, and P. Larsson-Edefors, "VLSI Implementations of Carrier Phase Recovery Algorithms for M-QAM Fiber-Optic Systems," *J. Lightwave Technol.*, vol. 38, no. 14, pp. 3616–3623, Jul. 2020, doi: 10.1109/JLT.2020.2976166.